\documentclass{article}
\usepackage{amssymb,amsmath,amsxtra,cite,bm}
\usepackage{graphicx}
\usepackage{float}
\usepackage[symbol]{footmisc}

\def\rd{\mathrm{d}}
\def\ri{\mathrm{i}}

\begin{document}
\begin{center}
{\large\textbf{Slipping and rolling on a rough accelerating surface}}
\\[\baselineskip]
{\sffamily M.~Khorrami$^{1,}\footnote{Corresponding Author E-mail: mamwad@alzahra.ac.ir}$, A.~Aghamohammadi$^1$,
C.~Aghamohammadi$^2$}
\\[\baselineskip]
$^1$ Department of Fundamental Physics, Faculty of Physics, Alzahra University, Tehran, Iran
\\[\baselineskip]
$^2$ Cold Spring Harbor Laboratory, Cold Spring Harbor, NY, USA
\end{center}

\vspace{2\baselineskip}
\begin{abstract}
\noindent The two-dimensional motion of an object on a moving
rough horizontal plane is investigated. Two cases are studied:
the plane having a translational acceleration, and a rotating plane.
For the first case, the motions of a point particle and a sphere
are studied, and it is shown that the solution to the latter problem
can be expressed in terms of the solution to the former one.
Examples of constant acceleration and periodic acceleration along a
fixed line, and specifically sinusoidal acceleration along a fixed line,
are studied in more detail. Also a situation is investigated where
the friction is anisotropic, that is the friction coefficient depends
on the direction of the velocity. In this situation, there may be
stick-slip motions, and these are investigated in detail. For the second
case, the motion of a point particle on a rough turntable is investigated.
\end{abstract}
\section{Introduction}
The one-dimensional motion of a sliding point particle on a rough horizontal
or an inclined plane is among the standard pedagogical problems in mechanics
\cite{Hall,KLep}. In \cite{Irodov,GHR}, the two-dimensional motion of a particle
sliding on an inclined plane has been studied for a special choice of the friction coefficient.
Although friction theories have been developed for centuries and numerous
situations have been studied in detail, but there are still many unanswered questions,
both in the framework of physics and mechanical engineering. Among these are
extensions to two-dimensions without any restriction on the geometry or
friction, replacing the particle with an extended object, and adding
an acceleration (translational or rotational) to the substrate.

The two-dimensional motion of a particle sliding on a rough inclined plane
has been investigated in \cite{CA}. In \cite{Aghamohammadi}, the motion of
a particle on an arbitrary surface but confined to a fixed vertical plane
has been investigated. The tangential motion at the contact of two solid objects
has been studied in \cite{FBUW}, where it was shown that the friction force and
the torque are inherently coupled. The example studied there, is a disk sliding
and spinning on a horizontal flat surface. It could make the problem easier
to study the motion (on a plane) of symmetric extended objects; for example
two-dimensional symmetric objects such as a disk \cite{ML2016,VE,JSA}, or a hoop
\cite{BO2019,JSA}; or three-dimensional symmetric objects such as a sphere, or
a cylinder \cite{Per2010,Cross}. Ref. \cite{Milne} is among the relatively
old  books in which some of such problems have been discussed. The following
and the references therein, provide some other examples.

The dynamics of a steel disc spinning on a horizontal rough surface has been
investigated both experimentally and theoretically in \cite{MLZZ2014}.
In \cite{AK1}, the two-dimensional motion of a generally non-circular non-uniform
cylinder on a flat horizontal surface has been addressed. In \cite{CA},
the equation of  motion for a sphere on a rough inclined plane
has been solved, using the solution of the equation of motion for
a point particle sliding on the same inclined plane with a different
friction coefficient. In \cite{AGJ}, the motion of a point particle sliding
on a turntable has been studied. There are some other sources on the motion
of a sphere rolling on a turntable. Examples are \cite{Weltner,Burns,Romer,GST}.

Some of these models have found applications as well. For example in
\cite{BaZo}, the results of \cite{CA} have been used for the study and analysis
of granular materials including the study of rotation phenomenon using particle methods.

Motions influenced by dry friction, can also result in the stick-slip phenomenon.
There are two modes in the stick-slip phenomenon: the stick mode occurs when
there is no relative motion between the two objects in contact, and the slip mode
occurs when there is a relative motion. This phenomenon is frequently seen
all around us. Examples are squeaking doors, and earthquakes taking place
during periods of rapid slip. Stick-slip phenomena can also produce
sounds through induced vibrations. Examples are drawing chalk on
a blackboard, or moving a violin bow. Stick-slip phenomena are also important
in mechanical engineering and many studies have been devoted to it. Examples are \cite{FAM2012,WHWM2020,FJ2014} and references therein.
Typically, in the slip-stick phenomenon for the friction between
the involved surfaces the coefficient of static friction is greater than
the coefficient of kinetic friction.
But in the situation studied here this is not the case.

Here a situation is studied that there are two solids directly in contact, without
an intermediate fluid lubrication layer. Friction is taken to be proportional to
the normal force, with the same value for static and dynamic coefficients.
Dry friction imposes the resistance in the relative motion of two objects
in contact.

More specifically, here the two-dimensional motion of an object on
an accelerating rough horizontal plane is investigated. The plane's
acceleration may be translational or rotational (produced by a turntable,
for example). This paper has essentially two parts. In the first part,
sections 2 and 3, the plane has a translation acceleration and the motion
of a point particle or a homogeneous sphere is studied. It is shown that the solution
to the latter problem can be expressed in terms of the solution to the former.
Examples of constant acceleration and periodic acceleration along a
fixed line, and specifically sinusoidal acceleration along a fixed line,
are studied in more detail. The motion on an inclined plane is equivalent
to the motion on a surface with constant acceleration. So the results obtained
here for constant acceleration are expected to be related to those of \cite{CA}.
But this does not apply to the results corresponding to a time-dependent acceleration.
In the case of sinusoidal acceleration along a fixed line, there may be
stick-slip motions, which are investigated in detail. Also a situation is
investigated where the friction is anisotropic, that is the friction coefficient
depends on the direction of the velocity. This leads to a much richer dynamic behavior.
The second part of the article, section 4, is on the motion of a point particle
on a rough turntable. Here the evolution equation is reduced to a set of
three first order coupled differential equations. The large-time
behavior of the system is studied in more detail, and some results are obtained
about the dependence of the large-time behavior on the initial condition;
specifically, which initial conditions result in final rest and which result
in perpetual motion. Finally, section 5 is devoted to the conclusion.
\section{A point particle sliding on a rough accelerating surface}
Consider a horizontal rough plane that is being pulled with the acceleration $\bm{A}$.
A particle of mass $m$ moves on this plane. There is friction between the particle
and the plane, with the coefficient $\mu$ which is assumed to be the same
for static and kinetic frictions. The equation of motion of the particle,
when it is not at rest, in the non-inertial frame of the accelerating plane is
\begin{align}\label{01}
\frac{\rd\,\bm{v}}{\rd\,t}&=-\bm{A}-\mu\,g\,\frac{\bm{v}}{v},
\intertext{where ${\bm{v}}$ is the particle's velocity, $g$ is
the gravity acceleration, and $|\bm{v}|$ is denoted by $v$. Multiplying this equation
by $(\bm{v}/v)$, one arrives at}
\label{2}
\frac{\rd\,v}{\rd\,t}&=-\bm{A}\cdot\frac{\bm{v}}{v}-\mu\,g.
\end{align}
This time evolution has some general properties:
\begin{itemize}
\item[$\bullet$] If the maximum of $|\bm{A}|$ is less than $(\mu\,g)$, the right-hand side
of (\ref{2}) is negative and the particle's speed decreases with time so that it will be
at rest in a finite time.
\item[$\bullet$] If for any time $t$ there is a larger time at which $|\bm{A}|$
is more than $(\mu\,g)$, then the particle cannot remain at rest indefinitely.
\item[$\bullet$] If the direction of $\bm{A}$ is fixed and $|\bm{A}|$ is always larger than
$(\mu\,g)$, then at sufficiently large times $v$ is increasing with time.
\item[$\bullet$] If the direction of $\bm{A}$ is fixed, then $|\bm{v}_\perp|$
decreases with time, where $\bm{v}_\perp$ is the part of $\bm{v}$ which is
perpendicular to $\bm{A}$. If $|\bm{A}|$ has an upper bound, then $\bm{v}_\perp$ tends
to zero at large times, so that the motion becomes essentially one dimensional.
\end{itemize}
\subsection{Constant acceleration}
A special case is when the acceleration is a constant vector.
One can choose the axes so that
\begin{align}\label{3}
\bm{A}&=-A\,\hat{\bm{x}},
\intertext{where $A$ is positive. Denoting the angle of $\bm{v}$ with the $x$ axis by $\theta$,}
\bm{v}&=v\,(\hat{\bm{x}}\,\cos\theta+\hat{\bm{y}}\,\sin\theta),
\intertext{where $(x,y)$ are Cartesian coordinates, and defining $\lambda$ through}
\lambda&=\frac{\mu\,g}{A},
\intertext{equation (\ref{01}) becomes}
\frac{\rd\,(v\,\cos\theta)}{\rd\,t}&=A\,(1-\lambda\,\cos\theta).\\ \label{07}
\frac{\rd\,(v\,\sin\theta)}{\rd\,t}&=-A\,\lambda\,\sin\theta.
\intertext{Eliminating $t$, one arrives at}
\frac{\rd\,(v\,\sin\theta)}{\rd\,(v\,\cos\theta)}&=
-\frac{\lambda\,\sin\theta}{1-\lambda\,\cos\theta}.
\intertext{So,}
\frac{\rd\,v}{v}&=\frac{\lambda-\cos\theta}{\sin\theta}\,\rd\,\theta,
\intertext{resulting in}
v\,(\sin\theta)\,\left(\tan\frac{\theta}{2}\right)^{-\lambda}&=\mbox{constant},
\end{align}
or,
\begin{equation}\label{11}
v=v_0\,\frac{\sin\theta_0}{\sin\theta}\,\left(\frac{\displaystyle{\tan\frac{\theta}{2}}}
{\displaystyle{\tan\frac{\theta_0}{2}}}\right)^\lambda.
\end{equation}
Putting this in equation
 (\ref{07}),
 \begin{equation}
A\,\rd\,t=-\frac{v_0\,\sin\theta_0}{\displaystyle{\left(\tan\frac{\theta_0}{2}\right)^\lambda}}\,
\frac{\rd\,\displaystyle{\left[\left(\tan\frac{\theta}{2}\right)^\lambda\right]}}{\lambda\,\sin\theta},
 \end{equation}
resulting in
\begin{equation}\label{13}
t=\frac{v_0\,(\lambda+\cos\theta_0)}{A\,(1-\lambda^2)}
\,\left[\frac{(\lambda +\cos\theta)\,\sin\theta_0}{(\lambda +\cos\theta_0)\,\sin\theta}\, \left(\frac{\displaystyle{\tan\frac{\theta}{2}}}{\displaystyle{\tan\frac{\theta_0}{2}}}\right)^\lambda
-1\right].
\end{equation}
Using (\ref{11}) and (\ref{13}), one arrives at
\begin{equation}
\label{14}
v=\frac{(1-\lambda^2)\,A\,t+v_0\,(\lambda+\cos\theta_0)}{(\lambda+\cos\theta)}.
\end{equation}

\begin{figure}
\centering
\includegraphics[width=0.95\linewidth]{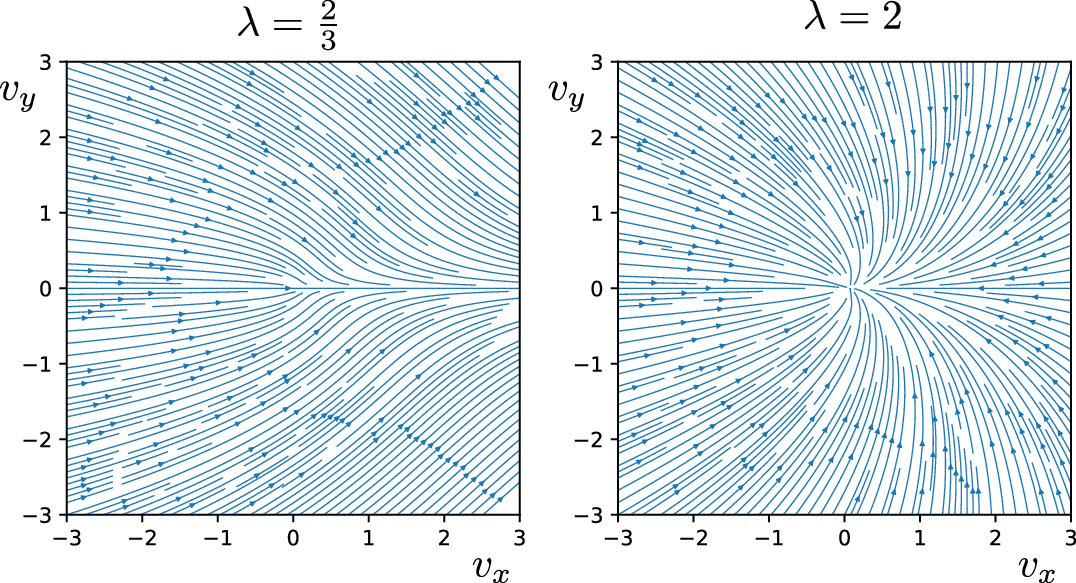}
\put(-245,-10){(a)}
\put(-75,-10){(b)}
\caption{The velocity phase portrait for different values of $\lambda$.
For $\lambda\leq 1$, the friction is not large enough to prevent the particle from sliding.
At large times, the particle slides along the plane's acceleration. For $\lambda> 1$,
the friction is large enough and the particle eventually comes to rest, in a finite time.}
\label{phase0}
\end{figure}

Depending on the value of $\lambda$, different behaviors occur:
\begin{itemize}
\item $\lambda=0$.\\
The surface is friction-less. The $y$-component of the particle's velocity is constant,
and the $x$-component of the particle's velocity grows linearly with time:
\begin{align}
v_x&=v_{0\,x}+A\,t.\\
v_y&=v_{0\,y}.
\end{align}
The particle's trajectory is generally a parabola.
\item $\lambda\leq 1$.\\
The friction is not large enough to keep the particle at rest. At large time, $\theta$
tends to zero, and the particle slides along the plane's acceleration.
For the special case $\lambda=1$,
\begin{align}
v\,(1+\cos\theta)&=v_0\,\,(1+\cos\theta_0).
\intertext{At large times the particle's velocity tends to a constant:}
v_x(\infty)&=\frac{v_0\,\,(1+\cos\theta_0)}{2}.\\
v_y(\infty)&=0.
\end{align}
\item $\lambda>1$.\\
The friction is large enough to prevent the particle from sliding. The particle
eventually stops at the time T:
\begin{align}\label{tvf3}
T=\frac{v_0\,(\lambda+\cos\theta_0)}{A\,(\lambda^2-1)}.
\end{align}
And as this time is approached, $\theta$ tends to zero, so the direction of
the particle's velocity approaches the $x$ axis.
\end{itemize}
These can be visualized in figure (\ref{phase0}), which is the phase portrait
of $(v_x,v_y)$ for different values of $\lambda$. For $\lambda\leq 1$, the friction
isn't large enough to prevent the particle from sliding: at large times, the particle slides
along $x$ direction (the plane's acceleration). For $\lambda> 1$, the friction is large enough
and the particle eventually comes to rest, after a finite time. At this point,
the direction of the velocity tends to the $x$ axis.
These results are consistent with those of ref. \cite{CA}.
\subsection{Sinusoidal linear acceleration}
\begin{figure}
\centering
\includegraphics[width=0.8\linewidth]{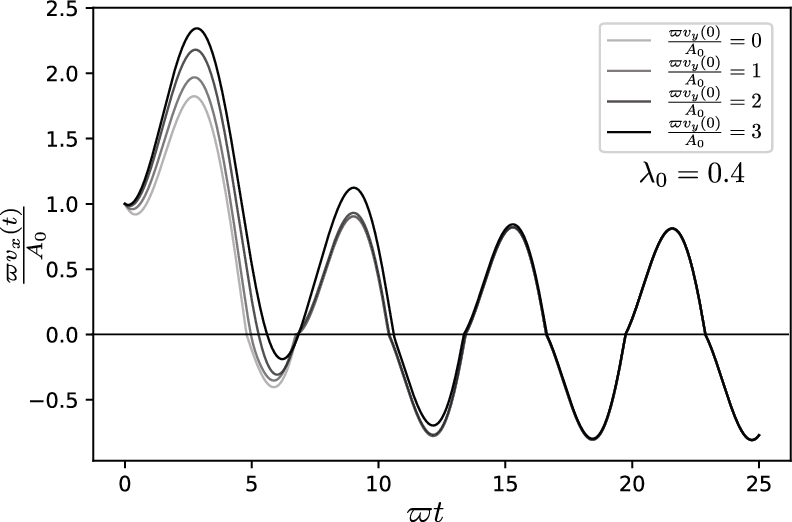}
\caption{At large times, the velocity becomes parallel to the vector $\bm{A}$ (here parallel
to the $x$ axis). The large time behavior of $v_x$ doesn't depend on the initial value of
$v_y$. But the transient behavior of $v_x$ does depend on the initial value of $v_y$. The figure shows
the time dependence of $v_x$ for a fixed initial value of $v_x$ and several initial values
of $v_y$.}
\label{phase02}
\end{figure}

Consider a case where the acceleration is sinusoidal and along a fixed line.
The axes are chosen so that
\begin{align}\label{21}
\bm{A}&=-A_0\,\sin(\varpi\,t)\,\hat{\bm{x}}.
\intertext{At large times, the $y$-component of the velocity vanishes.
If $A_0$ (which is taken to be positive) is smaller than $(\mu\,g)$,
$v_x$ vanishes at large times as well. Here the case is studied that
$A_0$ is larger than $(\mu\,g)$. So $v_x$ doesn't tend to zero
at large times. The dimensionless parameters $\lambda_0$, $\phi$, and $\mathfrak{w}$
are defined through}\label{22}
\lambda_0&=\frac{\mu\,g}{A_0}.\\ \label{23}
\phi&=\varpi\,t.\\ \label{24}
\mathfrak{w}&=\frac{\varpi\,v_x}{A_0}.
\intertext{The assumption that $A_0$ is larger than $(\mu\,g)$ means that}\label{25}
\lambda_0&<1.
\end{align}
At large times, the $v_y$ vanishes. If initially $v_y$
is much larger than the $v_x$, then initially the friction is approximately
along the $y$ axis, and is constant. That produces a constant acceleration
along the $y$ axis which decreases $v_y$ linearly with time.
But at some time $v_y$ becomes negligible compared to
$v_x$. After this, the friction is nearly along the $x$ axis,
with a small $y$-component which is proportional to $v_y$.
This results in a slower decrease of $v_y$, which is
exponential in time. Regarding $v_x$, its large time behavior
doesn't depend on the initial value of $v_y$. But its transient behavior does.
Figure \ref{phase02} shows this behavior for several initial values of $v_y$.
The large-time equation for $v_x$, when it is not zero, is rewritten as
\begin{align}
\label{26}
\frac{\rd\,\mathfrak{w}}{\rd\,\phi}&=\sin\phi-\lambda_0\,\frac{\mathfrak{w}}{|\mathfrak{w}|}.
\intertext{$|\sin\phi|$ becomes equal to $\lambda_0$ at four points
(in each period). For $\phi$ in $[0,2\,\pi]$, these points are denoted by $\phi_1$,
$\phi_2$, $\phi_3$, and $\phi_4$:}
\phi_1&=\sin^{-1}\lambda_0.\\
\phi_2&=\pi-\sin^{-1}\lambda_0.\\
\phi_3&=\pi+\sin^{-1}\lambda_0.\\
\phi_4&=2\,\pi-\sin^{-1}\lambda_0.
\end{align}
If there are time intervals in which $\mathfrak{w}$ is zero, then those time
intervals would end at $\phi_1$ or $\phi_3$ (in the first period). Assuming that
$\mathfrak{w}$ is zero at $\phi_1$, one arrives at the following expression for $\mathfrak{w}$,
for the interval in which $\mathfrak{w}$ is positive.
\begin{equation}
\mathfrak{w}=-\lambda_0\,(\phi-\phi_1)+\cos\phi_1-\cos\phi.
\end{equation}
This is valid for $\phi$ in $[\phi_1,\tilde\phi_2]$, where $\tilde\phi_2$
is the smallest value larger than $\phi_1$ which satisfies
\begin{equation}
-\lambda_0\,(\tilde\phi_2-\phi_1)+\cos\phi_1-\cos\tilde\phi_2=0.
\end{equation}
It is seen that $\tilde\phi_2$ is larger than $\phi_2$ but smaller than $\phi_4$.
If $\tilde\phi_2$ is smaller than $\phi_3$, then there is an interval $[\tilde\phi_2,\phi_3]$
in which $\mathfrak{w}$ is zero. Then $\mathfrak{w}$ is of the following form.
\begin{equation}
\mathfrak{w}=\begin{cases}
-\lambda_0\,(\phi-\phi_1)+\cos\phi_1-\cos\phi,&\phi_1<\phi<\tilde\phi_2\\
0,&\tilde\phi_2<\phi<\phi_3\\
\lambda_0\,(\phi-\phi_3)+\cos\phi_3-\cos\phi,&\phi_3<\phi<\tilde\phi_2+\pi\\
0,&\tilde\phi_2+\pi<\phi<\phi_1+2\,\pi
\end{cases}.
\end{equation}
And of course $\mathfrak{w}$ continues periodically in $\phi$, with the period $(2\,\pi)$.
For this behavior to happen, $\tilde\phi_2$ should be smaller than $\phi_3$, or
\begin{align}
-\lambda_0\,(\phi_3-\phi_1)+\cos\phi_1-\cos\phi_3&<0.
\intertext{Expressing everything in terms of $\lambda_0$, one arrives at
the following condition for $\lambda_0$:}
-\pi\,\lambda_0+2\,\sqrt{1-\lambda_0^2}&<0,
\end{align}
or,
\begin{align}\label{s16}
\lambda_0&>\lambda_\mathrm{c}.
\intertext{Where,}\label{s16-2}
\lambda_\mathrm{c}&=\frac{2}{\sqrt{4+\pi^2}}.
\end{align}

If the condition (\ref{s16}) is not satisfied, then the interval
$[\phi_1,\tilde\phi_2]$ in which the frictional acceleration is $(-\mu\,g)$ is
more than $\pi$, which is half the period. So whatever
the friction be in the interval $[\tilde\phi_2,\phi+2\,\pi]$,
the integral of the frictional acceleration from $\phi_1$ to
$(\phi_1+2\,\pi)$ is negative, and as the integral of $A$
in that interval is zero, $[\mathfrak{w}(\phi_1+2\,\pi)]$ would be negative.
This means that if  (\ref{s16}) is not satisfied, then there is
no periodic solution for $\mathfrak{w}$ which vanishes on some interval.
($\mathfrak{w}$ still does vanish at some points, but not on any interval.)
The friction is simply too small to do so. In that case,
the large-time behavior of $\mathfrak{w}$ is still periodic, but with the following form.
\begin{equation}
\mathfrak{w}=\begin{cases}-\lambda_0\,(\phi-\tilde\phi_1)+\cos\tilde\phi_1-\cos\phi,&
\tilde\phi_1<\phi<\tilde\phi_1+\pi\\
\lambda_0\,(\phi-\tilde\phi_1-\pi)-\cos\tilde\phi_1-\cos\phi,&
\tilde\phi_1+\pi<\phi<\tilde\phi_1+2\,\pi
\end{cases},
\end{equation}
where
\begin{align}
\tilde\phi_1&=\cos^{-1}\frac{\pi\,\lambda_0}{2}.
\intertext{And it is seen that when this behavior occurs, for which}
\lambda_0&<\lambda_\mathrm{c},
\intertext{then}
\tilde\phi_1&>\phi_1.
\end{align}

\begin{figure}[ht]
\centering
\includegraphics[width=0.5\linewidth]{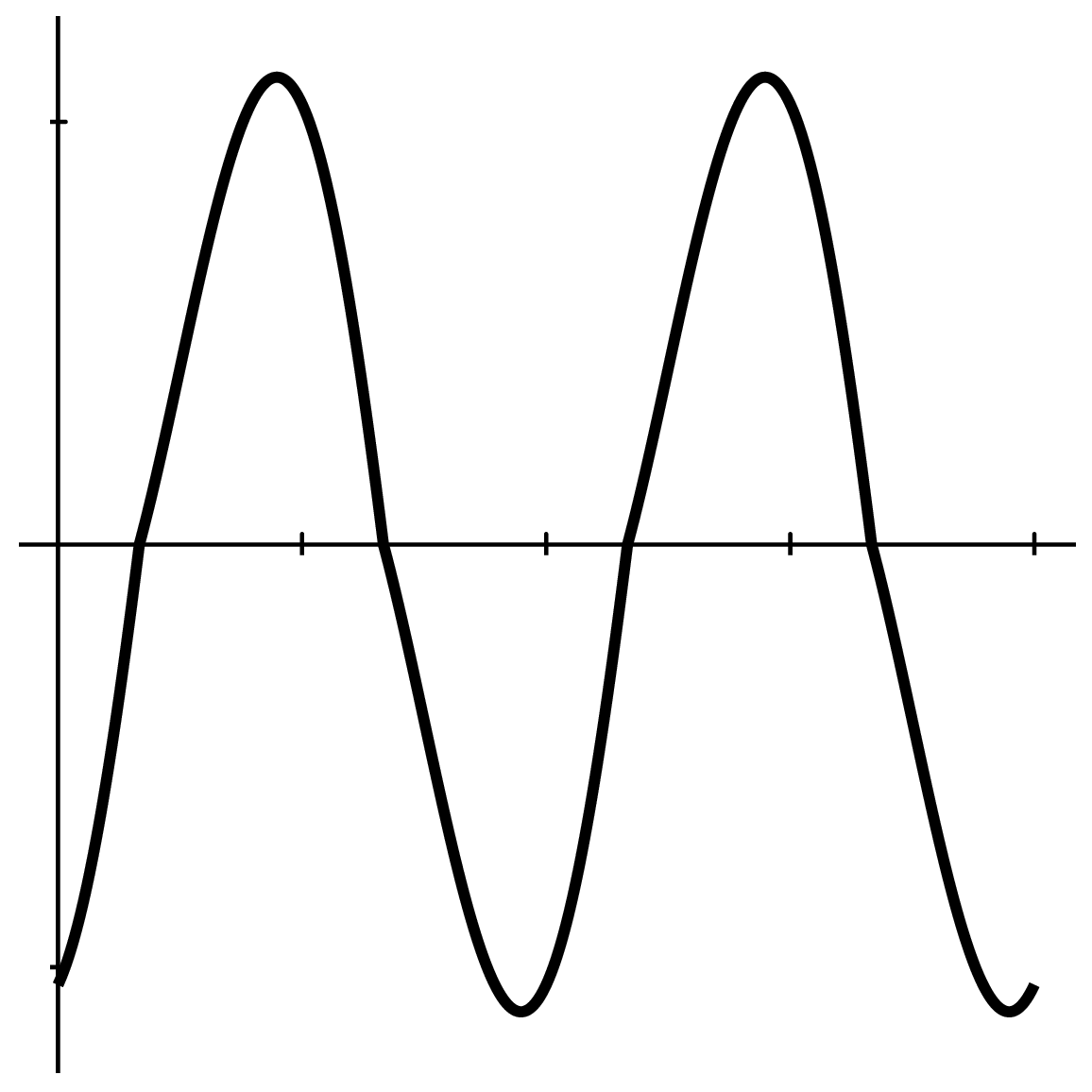}
\put(0,85){$\phi$}
\put(-167,173){$\mathfrak{w}$}
\put(-15,95){$4\,\pi$}
\put(-180,150){$0.8$}
{\linethickness{0.1pt}
\put(-165,170){\line(1,0){163}}
\put(-165,2){\line(1,0){163}}
\put(-2,2){\line(0,1){168}}
}
\\
\caption{\label{f1} The large-time behavior of $\mathfrak{w}$ (dimensionless $v_x$ or $u_x$) versus
$\phi$ (dimensionless $t$), for the small-friction case
$\lambda_0=\displaystyle{\frac{1}{\pi}}$.}
\end{figure}

\begin{figure}[ht]
\centering
\includegraphics[width=0.5\linewidth]{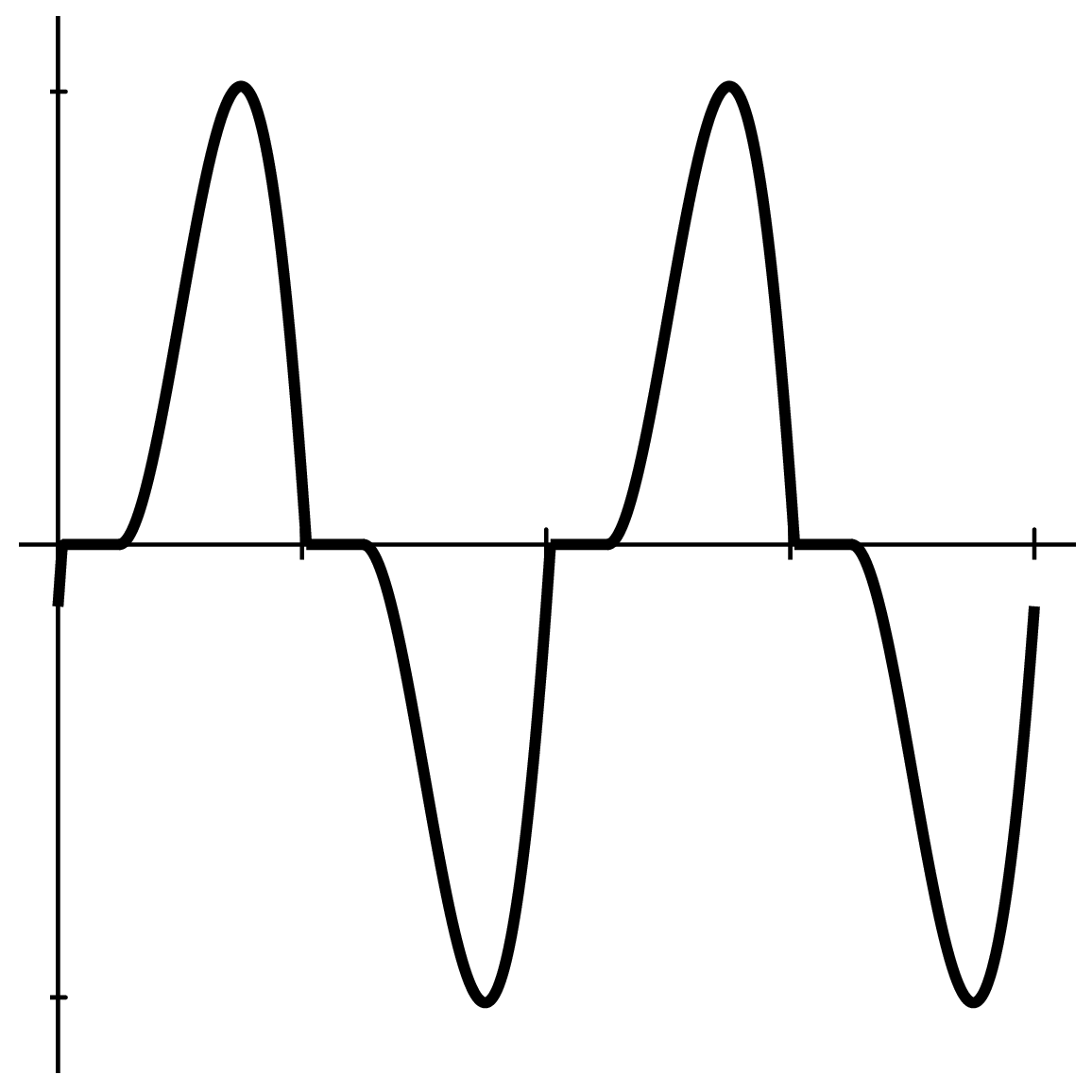}
\put(0,85){$\phi$}
\put(-167,173){$\mathfrak{w}$}
\put(-15,95){$4\,\pi$}
\put(-180,155){$0.3$}
{\linethickness{0.1pt}
\put(-165,170){\line(1,0){163}}
\put(-165,2){\line(1,0){163}}
\put(-2,2){\line(0,1){168}}
}
\\
\caption{\label{f2} The large-time behavior of $\mathfrak{w}$ (dimensionless $v_x$ or $u_x$) versus
$\phi$ (dimensionless $t$), for the large-friction case
$\lambda_0=\displaystyle{\frac{1}{\sqrt{2}}}$.}
\end{figure}
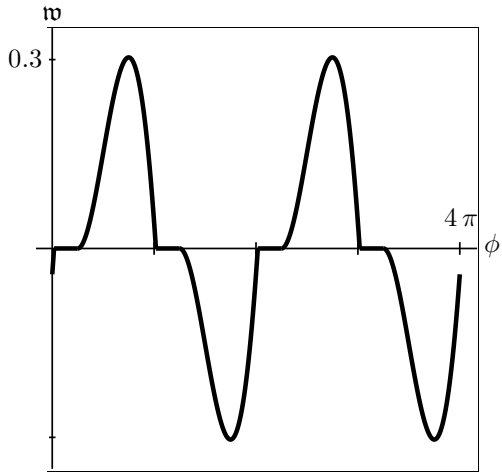

To summarize, the qualitative large-time behavior of $\mathfrak{w}$ (hence $v_x$)
depends on $\lambda_0$:
\begin{itemize}
\item[$\bullet$] $\lambda_0<\lambda_\mathrm{c}$.\\
In this small-friction situation, at each period $v_x$ vanishes at exactly two points.
\item[$\bullet$] $\lambda_\mathrm{c}<\lambda_0<1$.\\
In this large-friction situation, each period contains two time intervals when $v_x$ is zero.
\item[$\bullet$] $1<\lambda_0$.\\
In this very-large-friction situation, $v_x$ is identically zero.
\end{itemize}
Figures \ref{f1} and \ref{f2} show examples of large-time behavior
of $v_x$ versus time, respectively.

\subsection{More general periodic linear acceleration}
The arguments of subsection 2.2 can qualitatively be repeated
for the case the acceleration is periodic. Consider a function $h$ substituting
the sine in the right hand side of (\ref{26}). It is assumed that $h$ is periodic with
the period $(2\,\pi)$, and the maximum of $|h|$ is $1$. The aim is to study
the large-time behavior of $\mathfrak{w}$. If $\lambda_0$ is larger than $1$
(very large friction), the particle will always be at rest. For a generic
$h$, it is expected that when $\lambda_0$ is slightly less than $1$ (large friction)
the particle is at rest for a big fraction of time, and when $\lambda_0$ is very small
(small friction) the velocity doesn't vanish on some interval. So there should be
a $\lambda_\mathrm{c}$ determining the boundary of large and small friction.
The actual situation could be more complicated. There could be more that
one value of $\lambda$ at which such change-of-behaviors occur: There could be
cases when the number of zero-velocity intervals in each period changes.

As a simple no-so-generic example, consider
\begin{align}\label{42}
h(\phi)&=\begin{cases}
1,&0<\phi<\eta\,\pi\\
0,&\eta\,\pi<\phi<\pi
\end{cases}.\\ \label{43}
h(\pi+\phi)&=-h(\phi).
\end{align}
Where $\eta$ is a constant between $0$ and $1$.
It is seen that if $\lambda_0$ is less than $1$ but not very much less than $1$, then
\begin{align}
\mathfrak{w}&=\begin{cases}
(1-\lambda_0)\,\phi,&0<\phi<\eta\,\pi\\
(1-\lambda_0)\,\eta\,\pi-\lambda_0\,(\phi-\eta\,\pi),&\eta\,\pi<\phi
<(\eta\,\pi/\lambda_0)\\
0,&(\eta\,\pi/\lambda_0)<\phi<\pi
\end{cases}.\\
\mathfrak{w}(\pi+\phi)&=-\mathfrak{w}(\phi).
\end{align}
The condition that {\em $\lambda_0$ is not very much less than $1$}, is that
$\mathfrak{w}$ vanishes before the friction becomes nonzero again. That is, $\lambda_0$
being large than $\lambda_\mathrm{c}$, where
\begin{align}\label{46}
\lambda_\mathrm{c}&=\eta.
\end{align}
This is qualitatively similar to what was found in subsection 2.2.
It is seen that for $\eta=1$ (the acceleration never vanishes), there is
no large-friction phase. And for $\eta=0$ (no acceleration), there is
no small-friction phase.
\subsection{Anisotropic friction coefficient}
Up to now, it has been assumed that the friction coefficient is isotropic.
That is, $\mu$ is a constant, specifically independent of the direction of
$\bm{v}$. Here a situation is studied in which this is not the case.
The case of linear sinusoidal acceleration is reexamined. Again,
at large times only the component of the velocity along the acceleration
could be nonzero. Then, for the large-time behavior equations similar to
(\ref{21}) to (\ref{24}) are used, except that (\ref{22}) is substituted with
\begin{alignat}{2}
\lambda_+&=\frac{\mu_+\,g}{A_0}.\\
\lambda_-&=\frac{\mu_-\,g}{A_0}.
\intertext{where $\mu_+$ ($\mu_-$) corresponds to positive (negative) $v_x$.
$\lambda_+$ and $\lambda_-$ are reparametrized through}
\lambda_+&=\lambda_0.\\
\lambda_-&=q\,\lambda_0.
\intertext{Then the equation (\ref{26}) becomes}
\frac{\rd\,\mathfrak{w}}{\rd\,\phi}&=\sin\phi-\lambda_0,&\quad\mathfrak{w}&>0.\\
\frac{\rd\,\mathfrak{w}}{\rd\,\phi}&=\sin\phi+q\,\lambda_0,&\quad\mathfrak{w}&<0.
\end{alignat}
Without loss of generality, one could consider $q$ to be not less than $1$.
Then, arguments similar to those presented for the isotropic case
result in the following regions for the qualitative large-time behavior of $\mathfrak{w}$,
in terms of $(q,\lambda_0)$. The change of behavior in this parameter space occurs
on the following curves:
\begin{alignat}{2}
&\mbox{\textsf{\bfseries A}}&K=\frac{\pi\,q}{q+1},
\quad \frac{\sqrt{1-\lambda_0^2}}{\lambda_0}&=\frac{K}{\sin^2 K}-\cot K.\\
&\mbox{\textsf{\bfseries B}}&\quad
\sqrt{1-\lambda_0^2} +\sqrt{1-(q\,\lambda_0)^2}&=
\lambda_0\,[\pi + \sin^{-1}(q\,\lambda_0) - \sin^{-1}(\lambda_0)].\\
&\mbox{\textsf{\bfseries C}}&\quad \lambda_0&=\frac{1}{q}.\\
&\mbox{\textsf{\bfseries D}}&\quad \lambda_0&=1.
\end{alignat}
The parameter space is divided into the following regions
\begin{itemize}
\item[\textsf{\bfseries I}] Below \textsf{\bfseries A}\\
Each period contains an interval of positive speed and an interval of
negative speed, with no rest intervals.
\item[\textsf{\bfseries II}] Above \textsf{\bfseries A} and below \textsf{\bfseries B}
and \textsf{\bfseries C}\\
Each period contains an interval of positive speed, immediately followed
by an interval of negative speed, and then an interval of rest.
\item[\textsf{\bfseries III}] Between \textsf{\bfseries B} and \textsf{\bfseries C}\\
Each period contains an interval of positive speed, an interval of negative speed,
and two intervals of rest between these.
\item[\textsf{\bfseries IV}] Between \textsf{\bfseries C} and \textsf{\bfseries D}\\
Each period contains an interval of positive speed and an interval of rest.
\item[\textsf{\bfseries V}] Above \textsf{\bfseries D}\\
The particle is at rest.
\end{itemize}
The curves \textsf{\bfseries A} and \textsf{\bfseries B} have an intersection
at the point $X$, and the curves \textsf{\bfseries B} and \textsf{\bfseries C}
have an intersection at the point $Y$. Actually they are tangent to each other at $Y$:
\begin{align}
X&=(q=1,\lambda_0=\lambda_\mathrm{c}).\\
Y&=(q=4.603,\lambda_0=0.217)
\end{align}

\begin{figure}[H]
\centering
\includegraphics[width=0.5\linewidth]{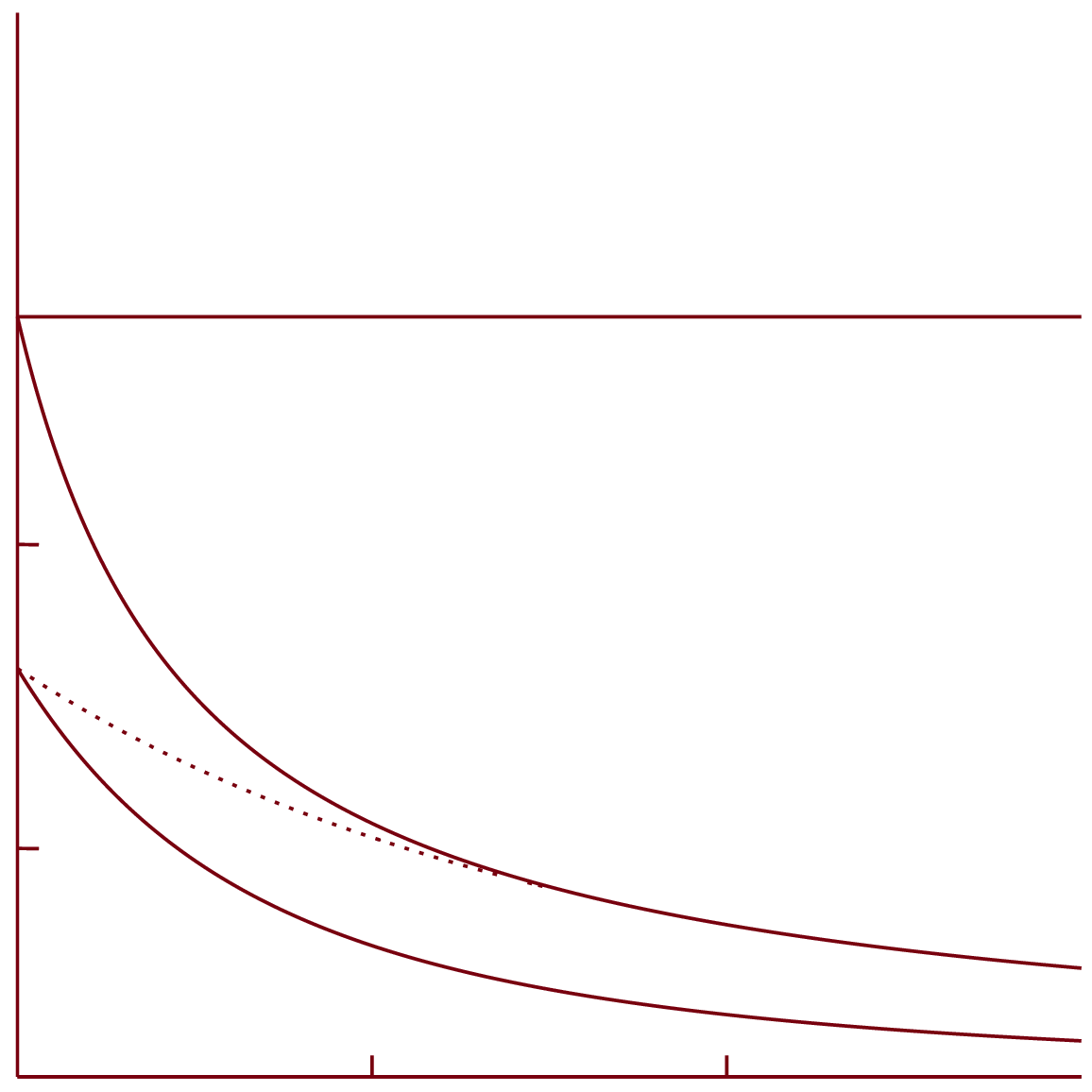}
\put(-7,-5){$q$}
\put(-182,163){$\lambda_0$}
\put(-185,122){$1$}
\put(-185,86){$0.7$}
\put(-185,38){$0.3$}
\put(-185,2){$0$}
\put(-172,-7){$1$}
\put(-116,-7){$3$}
\put(-59,-7){$5$}
\put(0,120){\textsf{\bfseries D}}
\put(0,15){\textsf{\bfseries C}}
\put(-155,53){\textsf{\bfseries B}}
\put(0,5){\textsf{\bfseries A}}
\put(-180,65){$X$}
\put(-172,65){$\scriptstyle{\bullet}$}
\put(-71,31){$Y$}
\put(-71,26.5){$\scriptstyle{\bullet}$}
\put(-150,15){\textsf{\bfseries I}}
\put(-130,33){\textsf{\bfseries II}}
\put(-165,70){\textsf{\bfseries III}}
\put(-100,70){\textsf{\bfseries IV}}
\put(-100,140){\textsf{\bfseries V}}
{\linethickness{0.1pt}
\put(-168,170){\line(1,0){166}}
\put(-2,2){\line(0,1){168}}
}
\\
\caption{\label{f3} The regions in the parameter space of the anisotropic
friction, corresponding to the large-time behavior of the velocity with
linear sinusoidal acceleration. The curve \textsf{\bfseries B} is the dotted curve.}
\end{figure}

\begin{figure}[H]
\begin{picture}(150,120)(0,0)
\includegraphics[scale=0.20]{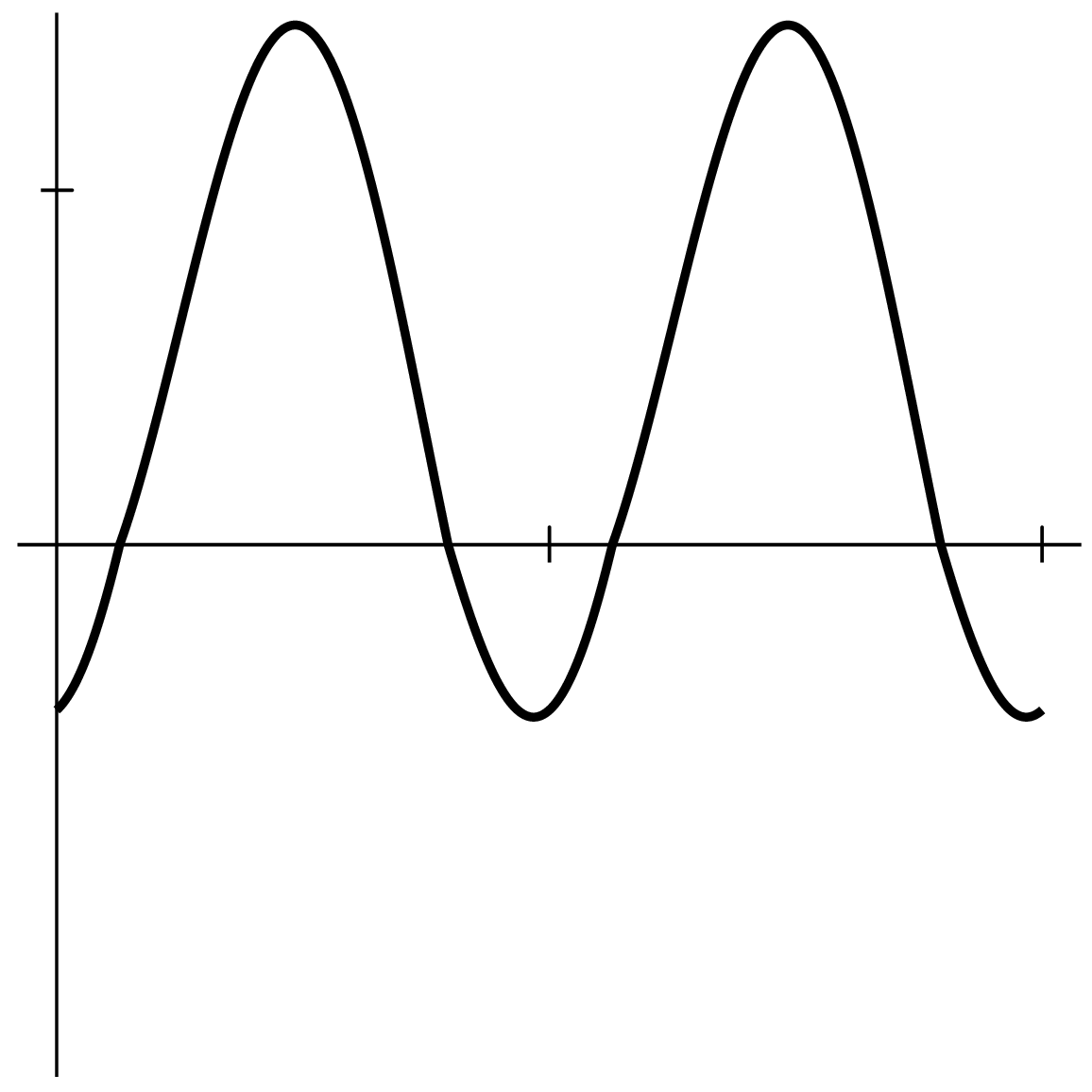}
\put(-110,113){$\mathfrak{w}$}
\put(-114,90){$1$}
\put(2,53){$\phi$}
\put(-11,60){$4\,\pi$}
\put(-57,25){\textsf{\bfseries I}}
{\linethickness{0.1pt}
\put(-105,110){\line(1,0){104}}
\put(-105,2){\line(1,0){104}}
\put(-2,2){\line(0,1){108}}
}
\end{picture}
\begin{picture}(150,120)(0,0)
\includegraphics[scale=0.20]{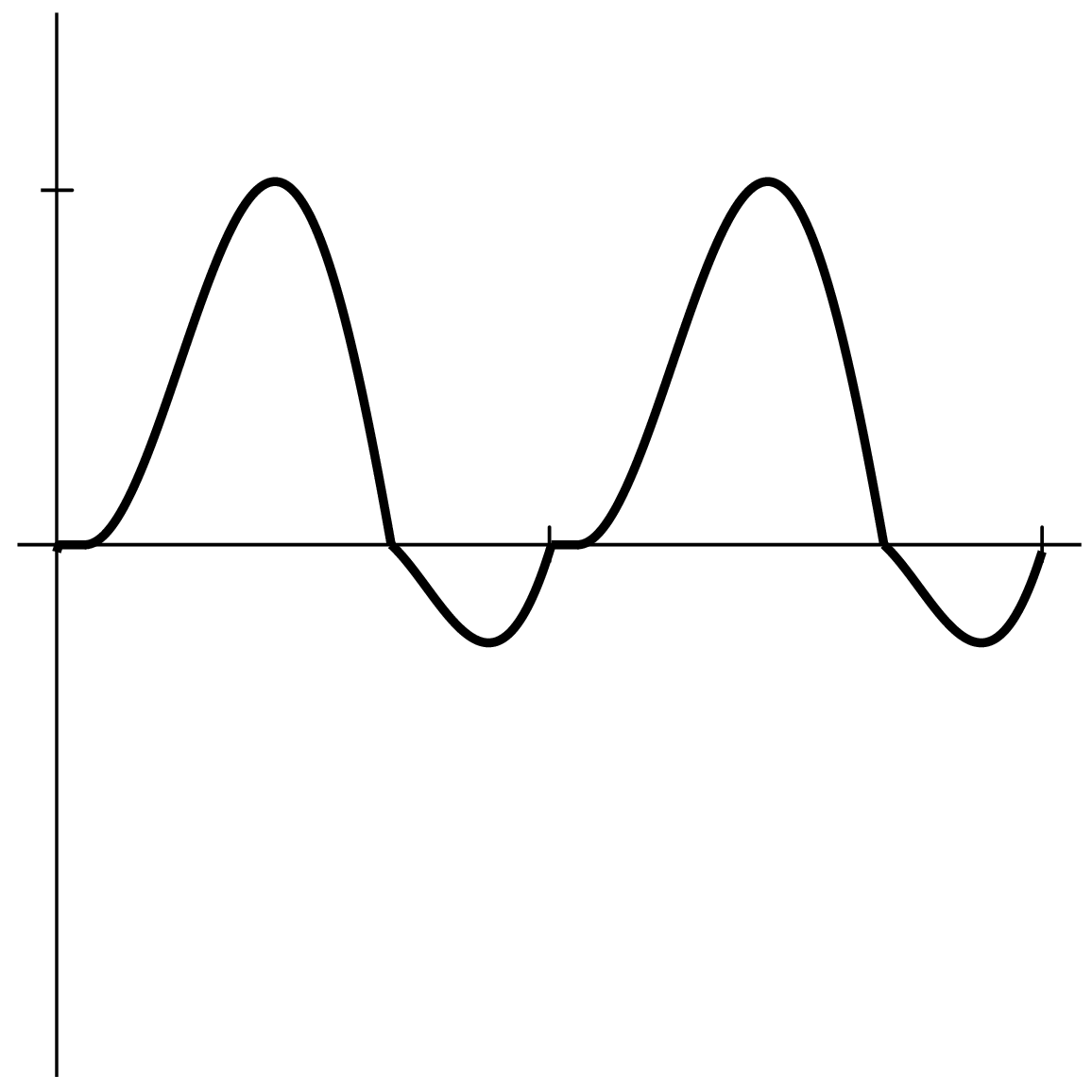}
\put(-110,113){$\mathfrak{w}$}
\put(-114,90){$1$}
\put(2,53){$\phi$}
\put(-11,60){$4\,\pi$}
\put(-59,25){\textsf{\bfseries II}}
{\linethickness{0.1pt}
\put(-105,110){\line(1,0){104}}
\put(-105,2){\line(1,0){104}}
\put(-2,2){\line(0,1){108}}
}
\end{picture}
\\[\baselineskip]
\begin{picture}(150,120)(0,0)
\includegraphics[scale=0.20]{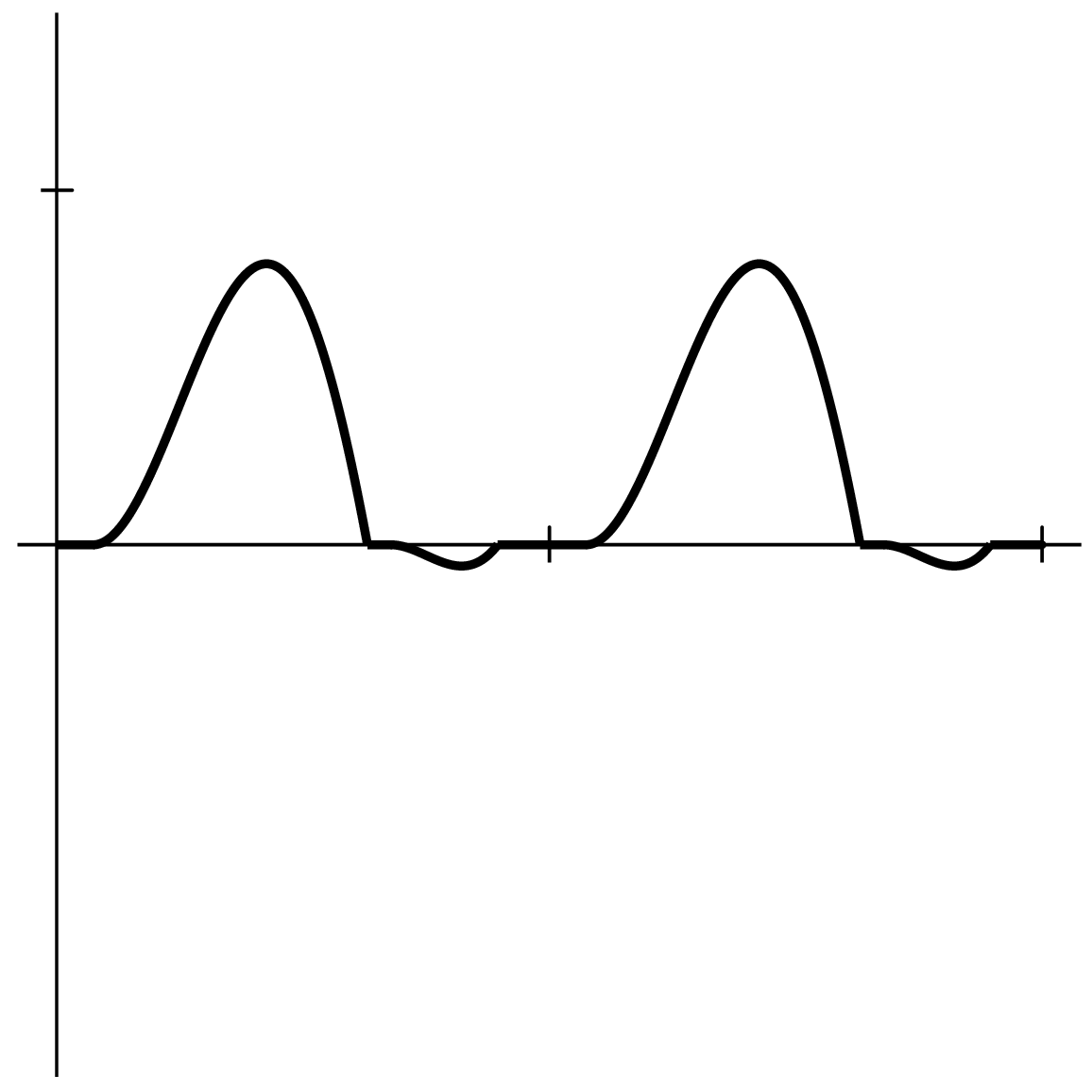}
\put(-110,113){$\mathfrak{w}$}
\put(-114,90){$1$}
\put(2,53){$\phi$}
\put(-11,60){$4\,\pi$}
\put(-60.3,25){\textsf{\bfseries III}}
{\linethickness{0.1pt}
\put(-105,110){\line(1,0){104}}
\put(-105,2){\line(1,0){104}}
\put(-2,2){\line(0,1){108}}
}
\end{picture}
\begin{picture}(150,120)(0,0)
\includegraphics[scale=0.20]{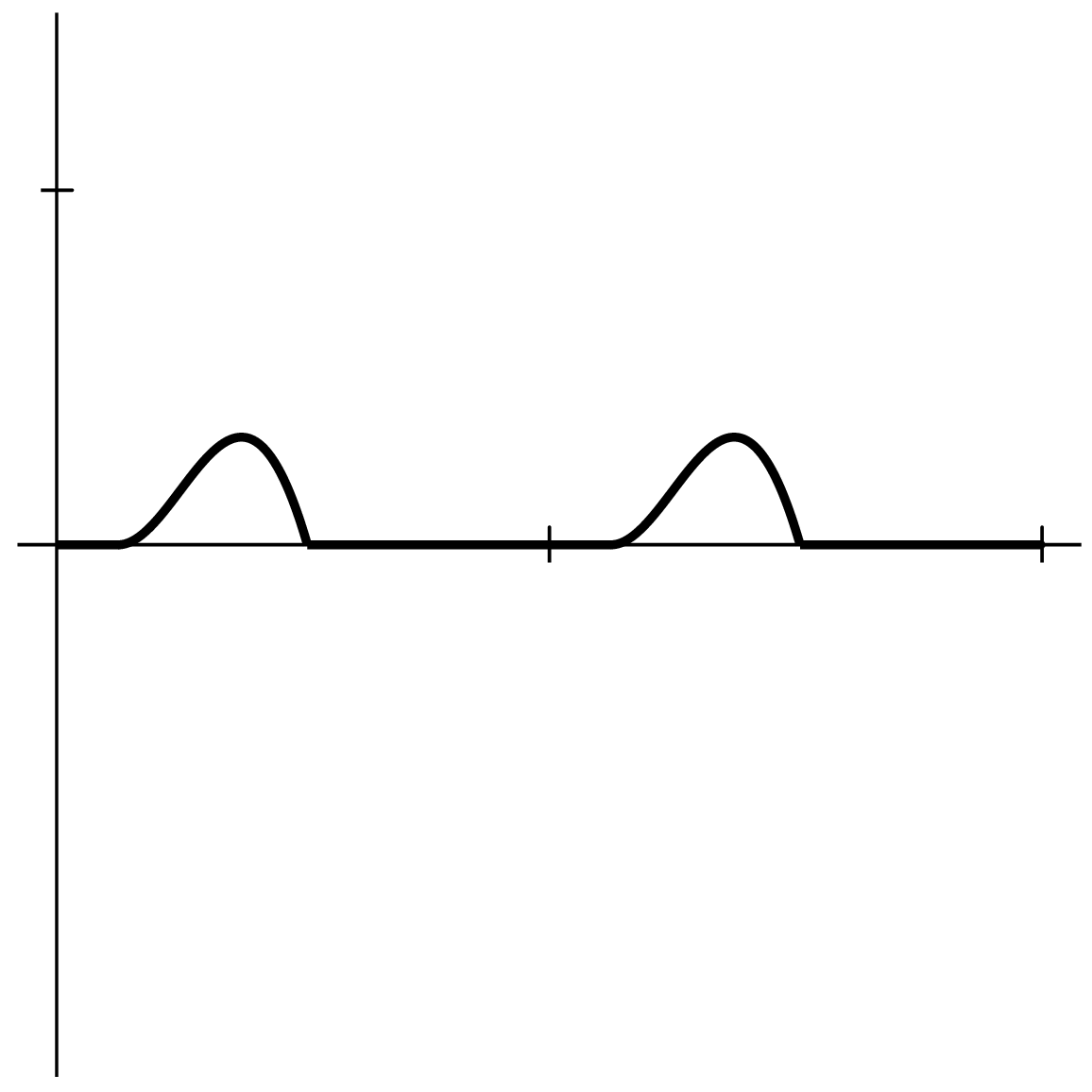}
\put(-110,113){$\mathfrak{w}$}
\put(-114,90){$1$}
\put(2,53){$\phi$}
\put(-11,60){$4\,\pi$}
\put(-60,25){\textsf{\bfseries IV}}
{\linethickness{0.1pt}
\put(-105,110){\line(1,0){104}}
\put(-105,2){\line(1,0){104}}
\put(-2,2){\line(0,1){108}}
}
\end{picture}
\\[\baselineskip]
\begin{picture}(150,120)(0,0)
\includegraphics[scale=0.20]{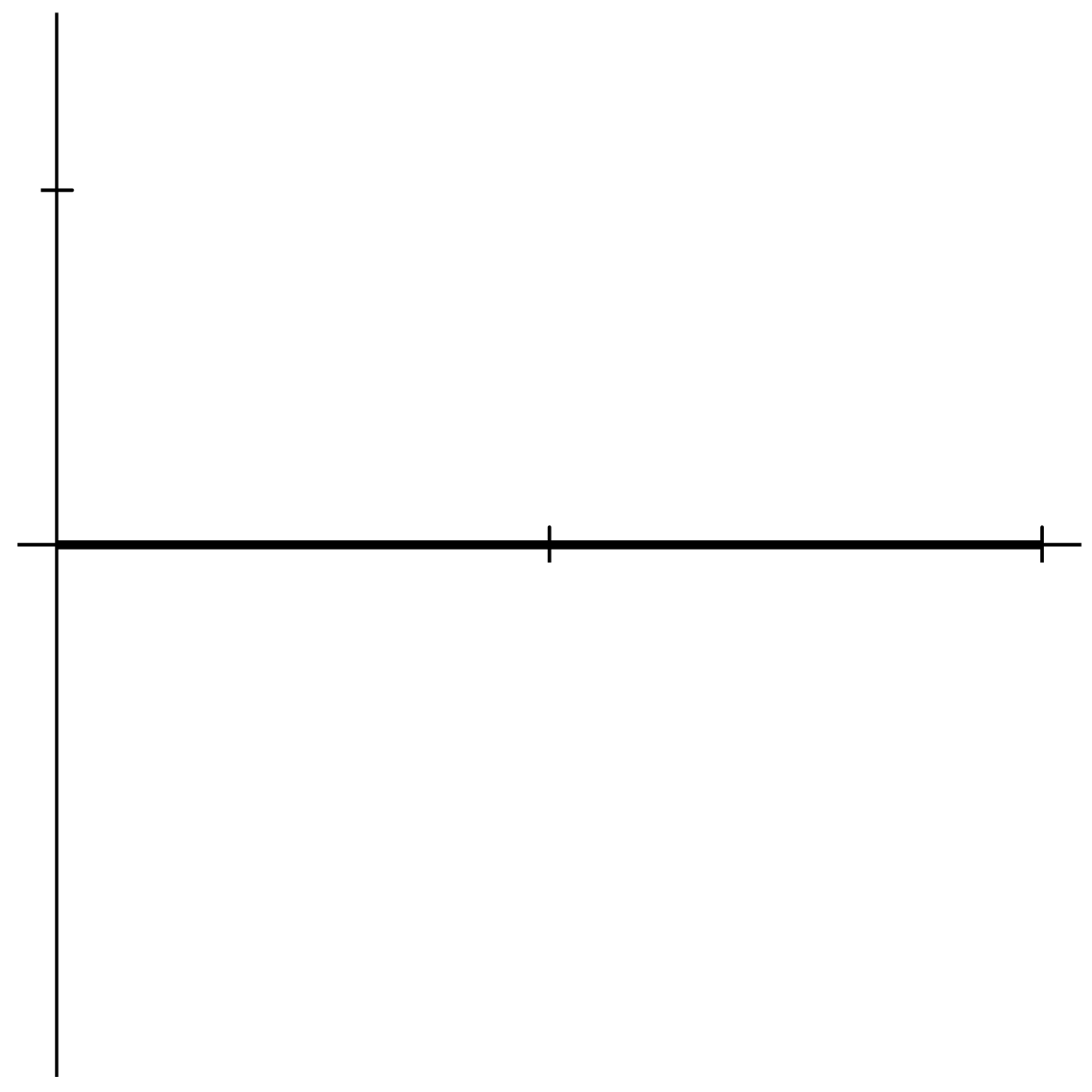}
\put(-110,113){$\mathfrak{w}$}
\put(-114,90){$1$}
\put(2,53){$\phi$}
\put(-11,60){$4\,\pi$}
\put(-59,25){\textsf{\bfseries V}}
{\linethickness{0.1pt}
\put(-105,110){\line(1,0){104}}
\put(-105,2){\line(1,0){104}}
\put(-2,2){\line(0,1){108}}
}
\end{picture}
\\
\caption{\label{f4} The large-time behavior of the velocity in each region
in the parameter space of the anisotropic friction, corresponding to
the large-time behavior of the velocity with linear sinusoidal acceleration.
Each plot contains two periods.}
\end{figure}
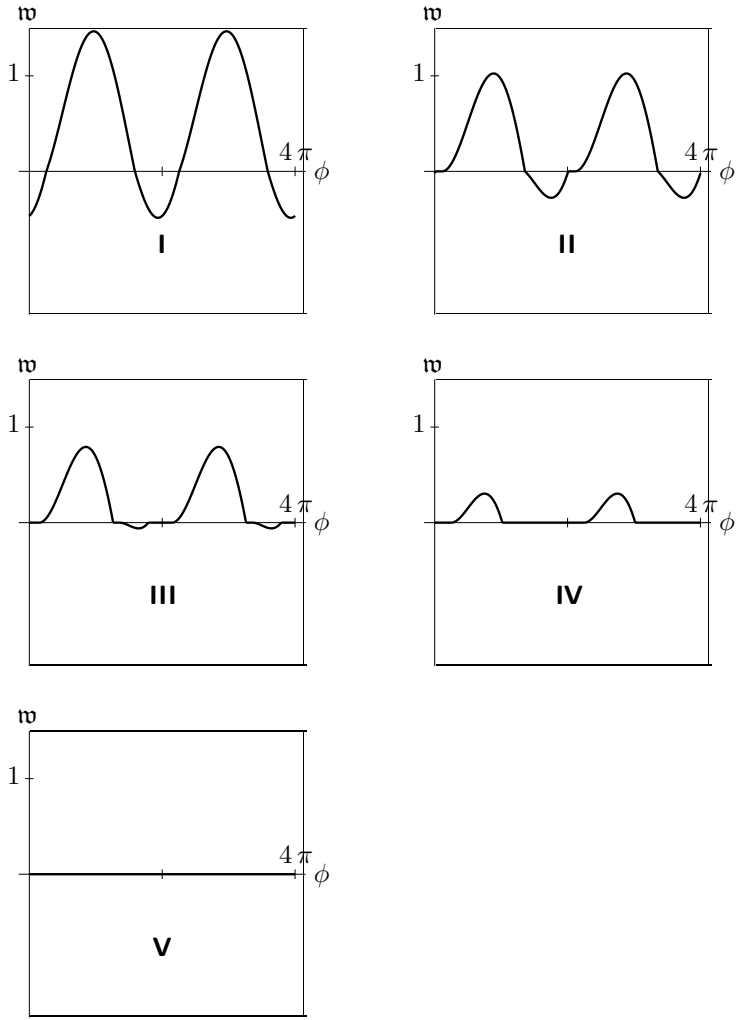

The regions of the parameter space are illustrated in the figure \ref{f3}.
The large-time behavior of the velocity in each region is illustrated in figure \ref{f4}.

\newpage
\vspace{3\baselineskip}

\section{The motion of a sphere on a rough accelerating horizontal surface}
In this section, the motion of a homogeneous sphere on a rough accelerating
horizontal plane is investigated. The sphere is of radius $R$ and mass $m$,
and it can roll and slide on the plane. The equation of motion for the sphere
(in the accelerated frame) are
\begin{align}\label{iplane18}
m\,\frac{\rd^2\,\bm{r}}{\rd\,t^2}&=-m\,\bm{A}+\bm{F},\\
I\,\frac{\rd\,\bm{\omega}}{\rd\,t}&=\bm{R}\times\bm{F},
\intertext{where $\bm{r}$ is the two-dimensional position of the center of
the sphere, $\bm{\omega}$ is the angular velocity of the sphere, $I$
is the moment of inertia of the sphere, $\bm{F}$ is the friction force, and}
\bm{R}&=-R\,\hat{\bm{z}}.
\intertext{The $z$ axis is normal to the plane and upward. Defining $\kappa$ through}
\kappa&=\frac{I}{m\,R^2},
\intertext{the equations of motion for the case the sphere slids become}
\frac{\rd^2\,\bm{r}}{\rd\,t^2}&=-\bm{A}-\mu\,g\,
\left|\frac{\rd\,\bm{r}}{\rd\,t}+\bm{\omega}\times\bm{R}\right|^{-1}\,
\left(\frac{\rd\,\bm{r}}{\rd\,t}+\bm{\omega}\times\bm{R}\right).\\
\kappa\,R^2\,\frac{\rd\,\bm{\omega}}{\rd\,t}&=-\mu\,g\,\bm{R}\times\left\{
\left|\frac{\rd\,\bm{r}}{\rd\,t}+\bm{\omega}\times\bm{R}\right|^{-1}\,
\left(\frac{\rd\,\bm{r}}{\rd\,t}+\bm{\omega}\times\bm{R}\right)\right\}.
\end{align}
So,
\begin{align}
\frac{\rd\,(\bm{\omega}\times\bm{R})}{\rd\,t}&=-\frac{\mu\,g}{\kappa}\,
\left|\frac{\rd\,\bm{r}}{\rd\,t}+\bm{\omega}\times\bm{R}\right|^{-1}\,
\left(\frac{\rd\,\bm{r}}{\rd\,t}+\bm{\omega}\times\bm{R}\right).\\
\frac{\rd}{\rd\,t}\,\left(\frac{\rd\,\bm{r}}{\rd\,t}+\bm{\omega}\times\bm{R}\right)
&=-\bm{A}-b\,\mu\,g\,
\left|\frac{\rd\,\bm{r}}{\rd\,t}+\bm{\omega}\times\bm{R}\right|^{-1}\,
\left(\frac{\rd\,\bm{r}}{\rd\,t}+\bm{\omega}\times\bm{R}\right),
\end{align}
where,
\begin{align}
b&=1+\frac{1}{\kappa}.
\end{align}
Using $R$ and $(\mu\,g)$, one can make the quantities dimensionless:
\begin{align}
\bm{\rho}&=\frac{\bm{r}}{R}.\\
\tau&=\sqrt{\frac{\mu\,g}{R}}\,t.\\
\bm{\zeta}&=\sqrt{\frac{R}{\mu\,g}}\,\bm{\omega}.\\
\bm{u}&=\frac{1}{\sqrt{\mu\,g\,R}}\,\left(\frac{\rd\,\bm{r}}{\rd\,t}+\bm{\omega}\times\bm{R}\right).\\
\bm{\gamma}&=\frac{\bm{A}}{\mu\,g}.
\end{align}
Denoting differentiation with respect to $\tau$ by dot, and denoting $|\bm{u}|$ by $u$,
the equations of motion become
\begin{align}\label{12}
\ddot{\bm{\rho}}&=-\bm{\gamma}-\frac{\bm{u}}{u}.\\ \label{35}
\dot{\bm{u}}&=-\bm{\gamma}-b\,\frac{\bm{u}}{u}.
\intertext{Of course one also has}
\bm{\zeta}&=\hat{\bm{z}}\,\zeta_3+\hat{\bm{z}}\times(\bm{u}-\dot{\bm{\rho}}).
\end{align}
And it is seen that $\zeta_3$ is a constant and does not enter the evolution
of other parameters.

Defining $\bm{\nu}$ as
\begin{align}
\dot{\bm{\nu}}&=\bm{\gamma},
\intertext{one arrives at}
(\dot{\bm{\rho}}+\bm{\nu})\spdot&=-\frac{\bm{u}}{u}.\\
(\bm{u}+\bm{\nu})\spdot&=-b\,\frac{\bm{u}}{u}.
\intertext{So,}\label{rhod}
\dot{\bm{\rho}}&=-\left(1-\frac{1}{b}\right)\,\bm{\nu}+\frac{\bm{u}}{b}+\bm{c},
\end{align}
where $\bm{c}$ is a constant vector. So the problem of finding the velocity and
the angular velocity of the sphere as a function of time, is reduced to finding $\bm{u}$
as a function of time. In other words to investigate the motion of a sphere on
a rough accelerating surface with the friction coefficient $\mu$ is equivalent
to study the motion of a particle  on a rough surface with the friction coefficient
$(b\,\mu)$ and the same acceleration.
\subsection{Constant acceleration}
Consider a special choice that $\bm{A}$ is a constant vector. The axes are chosen
like (\ref{3}). As noted before, the problem of finding $\bm{u}$
is the same as the problem of finding $\bm{v}$ for the motion of
a particle (without rotation), but with $\mu$ replaced by $(b\,\mu)$.
So one can use the results for $\bm{v}$ to obtain $\bm{u}$. Denoting the angle
of $\bm{u}$ with the $x$ axis by $\theta$, and defining $\lambda$ through
\begin{align}
\lambda&=\frac{b\,\mu\,g}{A},
\intertext{one arrives at}
u&=u_0\,\frac{\sin\theta_0}{\sin\theta}\,\left(\frac{\displaystyle{\tan\frac{\theta}{2}}}
{\displaystyle{\tan\frac{\theta_0}{2}}}\right)^\lambda.\\
\tau&=\frac{u_0\,(\lambda+\cos\theta_0)}{1-\lambda^2}
\,\left[\frac{(\lambda +\cos\theta)\,\sin\theta_0}{(\lambda +\cos\theta_0)\,\sin\theta}\, \left(\frac{\displaystyle{\tan\frac{\theta}{2}}}{\displaystyle{\tan\frac{\theta_0}{2}}}\right)^\lambda
-1\right].\\
u&=\frac{(1-\lambda^2)\,\tau+u_0\,(\lambda+\cos\theta_0)}{(\lambda+\cos\theta)}.
\end{align}
\subsection{Sinusoidal linear acceleration}
The (dimensionless) time evolution of $\bm{u}$ is the same as
the (dimensionless) time evolution of $\bm{v}$ as discussed in subsection 2.2, except that here
$\mu$ should be substituted with $(b\,\mu)$, so that here
\begin{equation}\label{72}
\lambda_0=\frac{b\,\mu\,g}{A_0}.
\end{equation}
Also, here a vanishing $\bm{u}$ means that the sphere rolls without slipping.
So the qualitative large-time behavior can be summarized as:
\begin{itemize}
\item[$\bullet$] $\lambda_0<\lambda_\mathrm{c}$.\\
In this small-friction situation, at each period $u_x$ vanishes at exactly two points.
At these points rolling occurs.
\item[$\bullet$] $\lambda_\mathrm{c}<\lambda_0<1$.\\
In this large-friction situation, each period contains two time intervals when $u_x$
is zero. At these intervals rolling occurs.
\item[$\bullet$] $1<\lambda_0$.\\
In this very-large-friction situation, $u_x$ is identically zero. This means that in this case,
eventually the motion of the sphere will be rolling without slipping.
\end{itemize}
Figures \ref{f1} and \ref{f2} show examples of large-time behavior
of the dimensionless $u_x$ versus time, respectively.
\subsection{More general periodic linear acceleration}
Similar to the subsection 3.2, one can bring the arguments of subsection 2.3
here, with $\lambda_0$ now defined like (\ref{72}). Also, for the example of
$h$ defined as (\ref{42}) and (\ref{43}), the boundary of small and large friction
is $\lambda_\mathrm{c}$, defined in (\ref{46}).
\subsection{Anisotropic friction coefficient}
Again, the (dimensionless) time evolution of $\bm{u}$ is the same as
the (dimensionless) time evolution of $\bm{v}$ as discussed in subsection 2.4,
except that here $\mu$ should be substituted with $(b\,\mu)$, so that here
\begin{equation}\label{85}
\lambda_0=\frac{b\,\mu_+\,g}{A_0}.
\end{equation}
\section{A point particle sliding on a rough turntable}
In this section the motion of a point particle on a turntable of infinite extent
is investigated. It is assumed that $\bm{\Omega}$, the angular frequency of
the rotation of the table, is perpendicular to the plane of
the table and is constant. The equation of motion in the rotating frame is
\begin{align}
\frac{\rd^2\,\bm{r}}{\rd\,t^2}&=\Omega^2\,\bm{r}+2\,\Omega\,\frac{\rd\,\bm{r}}{\rd\,t}\times\hat{\bm{z}}
-\mu\,g\,\left|\frac{\rd\,\bm{r}}{\rd\,t}\right|^{-1}\,\frac{\rd\,\bm{r}}{\rd\,t}.
\end{align}
From now on, it is assumed that $\Omega$ is positive. This is no loss of generality,
as any solution to the above equation with the sign of $\Omega$ changed,
is the mirror-reflected of a solution of the original problem. Using
$\Omega$ and $(\mu\,g)$, two dimensionless quantities $\bm{\rho}$ and $\tau$ are defined:
\begin{align}
\bm{\rho}&=\frac{\Omega^2}{\mu\,g}\,\bm{r}.\\
\tau&=\Omega\,t.
\intertext{The equation of motion becomes}
\ddot{\bm{\rho}}&=\bm{\rho}+2\,\dot{\bm{\rho}}\times\hat{\bm{z}}
-\frac{\dot{\bm{\rho}}}{|\dot{\bm{\rho}}|},
\end{align}
where dot means differentiation with respect to $\tau$. One has
\begin{align}
\dot{\bm{\rho}}\cdot\ddot{\bm{\rho}}&=\bm{\rho}\cdot\dot{\bm{\rho}}-|\dot{\bm{\rho}}|.\\
\hat{\bm{z}}\cdot\bm{\rho}\times\ddot{\bm{\rho}}&=-2\,\bm{\rho}\cdot\dot{\bm{\rho}}
-\frac{\hat{\bm{z}}\cdot\bm{\rho}\times\dot{\bm{\rho}}}{|\dot{\bm{\rho}}|}.\\
\bm{\rho}\cdot\ddot{\bm{\rho}}+\dot{\bm{\rho}}\cdot\dot{\bm{\rho}}&=\bm{\rho}\cdot\bm{\rho}
+2\,\hat{\bm{z}}\cdot\bm{\rho}\times\dot{\bm{\rho}}
-\frac{\bm{\rho}\cdot\dot{\bm{\rho}}}{|\dot{\bm{\rho}}|}+\dot{\bm{\rho}}\cdot\dot{\bm{\rho}}.
\end{align}
The length of $\dot{\bm{\rho}}$ is denoted by $p$:
\begin{align}
p&=|\dot{\bm{\rho}}|,
\intertext{and $\psi$ is defined as the angle of the position vector with respect
to the velocity vector, counterclockwise. Further, $\xi$ and $\ell$ are defined as}
\xi&=\rho\,\cos\psi.\\
\ell&=-\rho\,\sin\psi.
\intertext{So,}
p\,\ell&=\hat{\bm{z}}\cdot\bm{\rho}\times\dot{\bm{\rho}},\\
p\,\xi&=\bm{\rho}\cdot\dot{\bm{\rho}},
\intertext{Then, using}
\bm{\rho}\cdot\bm{\rho}&=\ell^2+\xi^2,
\intertext{one arrives at}\label{13-2}
\dot p&=\xi-1.\\ \label{14-2}
\dot\ell&=-\frac{\xi\,(2\,p+\ell)}{p}.\\ \label{15}
\dot\xi&=\frac{(p+\ell)^2}{p}.
\end{align}
Equation (\ref{13-2}) is the projection of Newton's equation
along the particle's velocity (or trajectory). Equation
(\ref{14-2}) is the projection of Newton's law along the azimuthal
direction (the equation of change for the angular momentum),
combined with the evolution equation for $u$. Equation (\ref{15})
is the projection of Newton's law along the radial direction,
combined with the evolution equation for $u$. Equations (\ref{13-2}),
(\ref{14-2}), and (\ref{15}) are three coupled differential equations,
governing the evolution of $p$, $\ell$, $\xi$. The system contains
a constant of motion. One has
\begin{align}
\rho\,\dot\rho&=\ell\,\dot\ell+\xi\,\dot\xi,\nonumber\\
&=\xi\,p,\nonumber\\
&=p\,(\dot p+1).
\intertext{So,}
\left(\frac{p^2-\rho^2}{2}\right)\spdot&=-p,
\intertext{or}\label{we}
\frac{p^2-\rho^2}{2}+s&=\mbox{constant},
\intertext{where $s$ is the arc-length parameter:}
\dot s&=p.
\end{align}
Re-dimensionalizing (\ref{we}), one arrives at an equation proportional to
\begin{equation}
\frac{m}{2}\,\left(\frac{\rd\,\bm{r}}{\rd\,t}\right)\cdot\left(\frac{\rd\,\bm{r}}{\rd\,t}\right)
-\frac{m\,\Omega^2\,r^2}{2}+m\,g\,s=\mbox{constant}.
\end{equation}
The first term is the kinetic energy of the particle in the non-inertial frame
of the turntable, the second term is the potential energy associated
to the centrifugal force, and $(-m\,g\,s)$ is the work
done by the friction. It is noted that the Coriolis force does no work.
So the above equation is the work-energy theorem in the non-inertial frame.

Expressing $\ell$ and $\xi$ in terms of $\rho$ and $\psi$, the evolution equations become
\begin{align}
\dot p&=\rho\,\cos\psi-1.\\
\dot\rho&=p\,\cos\psi.\\
\dot\psi&=2-\left(\frac{p}{\rho}+\frac{\rho}{p}\right)\,\sin\psi.
\intertext{One can also express $\rho$ and $p$ in terms of
hyperbolic parameters. There are two cases. Either,}
p^2-\rho^2&<0.\\
p&=A\,\sinh\alpha.\\
\rho&=A\,\cosh\alpha.
\intertext{Then,}
\dot A&=\sinh\alpha.\\
\dot\alpha&=\cos\psi-\frac{\cosh\alpha}{A}.\\
\dot\psi&=2\,[1-(\sin\psi)\,\coth(2\,\alpha)].
\intertext{Or,}
p^2-\rho^2&>0.\\
p&=B\,\cosh\beta.\\
\rho&=B\,\sinh\beta.
\intertext{Then,}
\dot B&=-\cosh\beta.\\
\dot\beta&=\cos\psi+\frac{\sinh\beta}{B}.\\
\dot\psi&=2\,[1-(\sin\psi)\,\coth(2\,\beta)].
\end{align}
The pair $(B,\beta)$ can be related to the pair $(A,\alpha)$ through
\begin{align}
A&=-\ri\,B.\\
\alpha&=\frac{\ri\,\pi}{2}+\beta.
\end{align}

Consider the first case (real $\alpha$). The evolution of $\psi$ has two
{\em quasi}-fixed points:
\begin{align}
\psi_1&=\sin^{-1}[\tanh(2\,\alpha)].\\
\psi_2&=\pi-\sin^{-1}[\tanh(2\,\alpha)].
\end{align}
These are not actual fixed points, as $\alpha$ is not constant. However,
$\psi_1$ is attractive while $\psi_2$ is repulsive. It is seen that $A$
increases indefinitely, unless $\alpha$ tends to zero. But if $\alpha$
is near zero, then $\psi$ changes rapidly towards its attractive
{\em quasi}-fixed point $\psi_1$, which is near zero, if $\alpha$ is
near zero. Then the evolution of $\alpha$ becomes
\begin{align}
\dot\alpha&\approx1-\frac{1}{A}.
\intertext{If}\label{115}
A&>1,
\end{align}
then $\alpha$ will increase and ceases to be near zero.
So the particle's speed will never vanish, and $A$ will
further increase. $A$ never decreases. So $A$ remains
bigger than $1$, if its initial value is bigger than $1$, and in that case $A$
increases indefinitely, and the particle's speed never vanishes.

Assuming that (\ref{115}) holds, initially, one can find the asymptotic behavior of
the variables at large times. Approximating $\psi$ by its attractive
{\em quasi}-fixed point, one arrives at
\begin{equation}
\dot\alpha\approx\frac{1}{\cosh(2\,\alpha)}-\frac{\cosh\alpha}{A}.
\end{equation}
$\alpha$ cannot remain finite, because in this case the right-hand side
	will eventually become positive (as $A$ increases). So both $A$ and $\alpha$ should
	increase indefinitely. A further approximation is to take $\alpha$ to be
	the {\em quasi}-fixed point of its evolution:
\begin{equation}
(\cosh\alpha)\,\cosh(2\,\alpha)\approx A.
\end{equation}	
For large values of $A$, and hence $\alpha$, this becomes
\begin{equation}
\alpha\approx\frac{1}{3}\,\ln(4\,A).
\end{equation}
So,
\begin{equation}
\dot A\approx\frac{(4\,A)^{1/3}}{2},
\end{equation}
leading to
\begin{align}
A&\sim\frac{2\,\tau^{3/2}}{\sqrt{27}}.\\
\alpha&\sim\frac{1}{2}\,\ln\frac{4\,\tau}{3}.\\
\psi&\sim\frac{\pi}{2}-\left(\frac{3}{\tau}\right)^{1/2}.\\
p&\sim\frac{2\,\tau^2}{9}.\\
\rho&\sim\frac{2\,\tau^2}{9}.\\
s&\sim\frac{2\,\tau^3}{27}.
\end{align}
As a check, it is seen that
\begin{equation}
\frac{A^2}{2}\sim s.
\end{equation}

Now consider cases when the particle eventually comes to rest.
The condition that the particle remain at rest, at $\rho=\sigma$, is
\begin{equation}\label{54}
\sigma\leq 1.
\end{equation}
The reason is that at $\rho>1$, the maximum of the static frictional force
is less than the centrifugal force. When the particle nears its rest, $p$ tends to zero,
so $\alpha$ tends to zero. Then from the evolution equation for $\psi$ it is seen that
$\psi$ tends to zero as well. So near the rest and before that,
\begin{align}
\dot\alpha&=1-\frac{1}{\sigma}+\cdots,
\intertext{which results in}
\alpha&=\frac{(\sigma-1)\,\tau}{\sigma}+\cdots.
\end{align}
Putting this in the evolution equation for $\psi$, and noting that
$\alpha$ and $\psi$ are both small, one arrives at
\begin{align}
\frac{\sigma-1}{\sigma}\,\frac{\rd\,\psi}{\rd\,\alpha}&
\approx 2\,\left(1-\frac{\psi}{2\,\alpha}\right),
\intertext{which results in}\label{58}
\psi&=2\,\alpha\,\left[\frac{\sigma}{2\,\sigma-1}+
c\,\alpha^{(2\,\sigma-1)/(1-\sigma)}\right]+\cdots,
\intertext{where $c$ is an integration constant. Similarly,
the evolution equation for $A$ becomes}
\frac{\sigma-1}{\sigma}\,\frac{\rd\,A}{\rd\,\alpha}&\approx\alpha,
\intertext{resulting in}\label{60}
A&=\sigma+\frac{\sigma}{\sigma-1}\,\frac{\alpha^2}{2}+\cdots.
\end{align}

For each value of $\sigma$, equations (\ref{58}) and (\ref{60})
represent a surface $\mathbb{S}_\sigma$ in the $3$-dimensional parameter space
$(A,\alpha,\psi)$. The envelope $\mathbb{S}$ of these surfaces is the boundary between
two regions of the parameter space: the region $\mathbb{V}_0$ corresponding
to the initial values which result in an eventual rest, and the region $\mathbb{V}_1$
corresponding to the initial values which result in unbounded motions.
To obtain the equation of $\mathbb{S}$, one notices that the equation of
$\mathbb{S}_\sigma$ is just (\ref{60}), with $\psi$ being free. Hence the parametric
equation for the envelope consists of (\ref{60}) and its derivative with respect to
the parameter $\sigma$:
\begin{align}
0&=1-\frac{1}{(\sigma-1)^2}\,\frac{\alpha^2}{2}+\cdots.
\intertext{Eliminating $\sigma$ between this and (\ref{60}), the equation for
$\mathbb{S}$ is determined as}
A&=\left(1-\frac{\alpha}{\sqrt{2}}\right)^2+\cdots.
\end{align}
The particle eventually comes to rest, if initially the value of $A$ is less
that the right-hand side. Otherwise its motion would be unbounded.
\section{Conclusion}
Even though friction-based problems in classical mechanics have been
studied extensively, there is still a substantial number of new researchs,
with compelling results. The problems stadied here were
the two-dimensional motion of a particle and a homogeneous sphere
on a moving rough horizontal plane.
It was shown that the problem of the motion of a homogeneous sphere on
a moving plane with translational acceleration is reduced to that of
the motion of a point particle.
Some examples were studied in more detail: constant acceleration, periodic (and
specifically sinusoidal) acceleration along a fixed line, and a situation where the
friction is anisotropic, which leads to a much richer dynamical behavior.
The motion of a point particle on a rough turntable was also studied.
It was shown that the evolution equation is reduced to a set of
three coupled first order differential equations.
The large-time behavior of the system was studied in more
detail, and some results were obtained about the dependence of
the large-time behavior on the initial condition; specifically,
which initial conditions result in final rest and which result in perpetual motion.
\\[\baselineskip]
\textbf{Acknowledgment}: The work of M. Khorrami and A. Aghamohammadi was supported by the
research council of the Alzahra University.


\newpage
\begin{thebibliography}{5}
\bibitem{Hall}          D. Halliday, R. Resnick, \& J. Walker; \textit{Fundamentals of physics}
                        (John Wiley \& Sons, extended ninth edition, 2010).
\bibitem{KLep}          D. Kleppner \& R. J. Kolenko; \textit{An introduction to mechanics}
                        (Cambridge University Press, 2010).
\bibitem{Irodov}        I. E. Irodov; \textit{Fundamental laws of mechanics} (Mir, 2002).
\bibitem{GHR}           P. Gn\"{a}dig, G. Honiyek, \& K. F. Riley;
                        \textit{200 Puzzling physics problems} (Cambridge University Press, 2001).
\bibitem{CA}            C. Aghamohammadi \& A. Aghamohammadi; \textit{Eur. J. Phys.} \textbf{32},
                        1049 (2011).
\bibitem{Aghamohammadi} A. Aghamohammadi; \textit{Eur. J. Phys.} \textbf{33}, 1111 (2012).
\bibitem{FBUW}          Z. Farkas, G. Bartels, T. Unger, \& D. E. Wolf; \textit{Phys. Rev. Lett.}
                        \textbf{90}, 248302 (2003).
\bibitem{ML2016}        D. Ma \& C. Liu; \textit{J. Appl. Mech.} \textbf{83}, 061003 (2016).
\bibitem{VE}            K. Voyenli \& E. Eriksen; \textit{Am. J. Phys.} \textbf{53}, 1149 (1985).
\bibitem{JSA}           M. A. Jalali, M. S. Sarebangholi, \& M. R. Alam; \textit{Phys. Rev.}
                        \textbf{E92}, 032913 (2015).
\bibitem{BAK}           A. V. Borisov,  A. A. Kilin, \& Y. L. Karavaev; \textit{PHYS-USP}
                        \textbf{60}, 931 (2017).
\bibitem{BO2019}        A. Bronars, \& O. M. \'OReilly; \textit{Proc. R. Soc. A} \textbf{475},
                        20190440 (2019).
\bibitem{Per2010}       B. N. J. Persson; \textit{Eur. Phys. J. E} \textbf{33}, 327 (2010).
\bibitem{Cross}         R. Cross; \textit{Eur. J. Phys.} \textbf{36}, 055011 (2015).
\bibitem{Milne}         E. A. Milne; \textit{Vectorial mechanics} (Methuen \& Co. Ltd., 1948).
\bibitem{MLZZ2014}      D. Ma,  C. Liu, Z. Zhao, \& H. Zhang; \textit{Proc. R. Soc. A} \textbf{470},
                        20140191 (2014).
\bibitem{AK1}           A. Aghamohammadi \& M. Khorrami; \textit{Can. J. Phys.} \textbf{96}, 627
                        (2018).
\bibitem{AGJ}           A. Agha, S. Gupta, \& T. Joseph; \textit{Am. J. Phys.} \textbf{83}, 126 (2015).
\bibitem{Weltner}       K. Weltner; \textit{Am. J. Phys.} \textbf{47}, 984 (1979).
\bibitem{Burns}         J. A. Burns; \textit{Am. J. Phys.} \textbf{49}, 56 (1981).
\bibitem{Romer}         R. H. Romer; \textit{Am. J. Phys.} \textbf{49}, 985 (1981).
\bibitem{GST}           J. Gersten, H. Soodak, \& M. S. Tiersten; \textit{Am. J. Phys.} \textbf{60},
                        43 (1992).
\bibitem{BaZo}          A. A. Bandeira \& T. I. Zohdi; \textit{Comput. Part. Mech.} \textbf{6}, 97 (2019).
\bibitem{FJ2014}        H. Fang \& J. Xu; \textit{Journal of Applied Mechanics} \textbf{81}, 051001 (2014).
\bibitem{FAM2012}       F. Mar\'{i}n, F. Alhama, \& J. A. Moreno; \textit{Int. J. Eng. Sci.} \textbf{60},
                        13, (2012).
\bibitem{WHWM2020}      X. C. Wang, B. Huang, R. L. Wang, J. L. Mo, \& H. Ouyang; \textit{MSSP}
                        \textbf{142}, 106705 (2020).
\end{thebibliography}
\end{document}